\documentclass[useAMS,usenatbib]{mn2e}
\usepackage{graphicx,amsmath,multirow,amssymb}
\usepackage{natbib}
\newcommand{\comment}[1]{}

\newcommand\Msun{M_{\odot}}
\def\simgt{\lower.5ex\hbox{$\; \buildrel > \over \sim \;$}}
\def\simlt{\lower.5ex\hbox{$\; \buildrel < \over \sim \;$}}

\title{Evolved stars in the Local Group galaxies. I. AGB evolution and dust production in IC 1613}
\author[Dell'Agli et al.]{F. Dell'Agli$^{1}$, M. Di Criscienzo$^{1}$, M.~L. Boyer$^{2,3}$, D. A. Garc\'{\i}a-Hern\'andez$^{4,5}$\\
$^1$INAF -- Osservatorio Astronomico di Roma, Via Frascati 33, 00040, Monte Porzio Catone (RM), Italy \\
$^2$CRESST and Observational Cosmology Lab, Code 665, NASA Goddard Space Flight Center, Greenbelt, MD 20771, USA\\
$^{3}$Department of Astronomy, University of Maryland, College Park, MD 20742 USA \\
$^{4}$Instituto de Astrof\'{\i}sica de Canarias, V\'{\i}a L\'actea s/n, E-38200 La Laguna, Tenerife, Spain \\
$^{5}$Departamento de Astrof\'{\i}sica, Universidad de La Laguna (ULL), E-38206 La Laguna, Tenerife, Spain\\
}

\begin{document}

\date{Accepted, Received; in original form }

\pagerange{\pageref{firstpage}--\pageref{lastpage}} \pubyear{2012}

\maketitle

\label{firstpage}

\begin{abstract}
We used models of thermally-pulsing asymptotic giant branch (AGB)
stars, that also describe the dust-formation process in the wind, to
interpret the combination of near-- and mid--infrared photometric data
of the dwarf galaxy IC 1613. This is the first time that this approach
is extended to an environment different from the Milky Way and the
Magellanic Clouds (MCs). Our analysis, based on synthetic population
techniques, shows a nice agreement between the observations and the
expected distribution of stars in the colour--magnitude diagrams
obtained with $JHK$ and {\it Spitzer} bands. This allows a
characterization of the individual stars in the AGB sample in terms of
mass, chemical composition, and formation epoch of the progenitors. We
identify the stars exhibiting the largest degree of obscuration as
carbon stars evolving through the final AGB phases, descending from
$1-1.25M_{\odot}$ objects of metallicity $Z=10^{-3}$ and from
$1.5-2.5M_{\odot}$ stars with $Z=2\times10^{-3}$. Oxygen--rich stars
constitute the majority of the sample ($\sim 65\%)$, mainly low mass
stars ($< 2M_{\odot}$) that produce a negligible amount of dust
($\leq10^{-7}M_{\odot}/{\rm yr}$). We predict the overall
dust-production rate from IC 1613, mostly determined by carbon stars, to be $\sim
  6\times10^{-7}M_{\odot}/{\rm yr}$ with an uncertainty of 30\%. The
capability of the current generation of models to interpret the AGB
population in an environment different from the MCs opens the
possibility to extend this kind of analysis to other Local Group
galaxies.
\end{abstract}


\section{Introduction}
Stars of mass in the range $1~ \Msun \leq M \leq 8~\Msun$, after the
consumption of helium in the core, evolve through the
thermally-pulsing asymptotic giant branch (AGB) phase. This
evolutionary phase, though extremely short compared to the duration of
the core hydrogen and helium burning, is of paramount importance,
because it is during the AGB phase that most of the mass loss occurs,
thus allowing the pollution of the interstellar medium with gas and
dust.

The importance of this class of objects stems from their relevance in
several astrophysical contexts, such as their effect on the
determination of the masses and star--formation rates of low-- and
high-- redshift galaxies \citep{maraston06,conroy09}, their role in
the chemical evolution of galaxies \citep{romano10}, and
likely the formation of second generation stars in globular
clusters \citep{ventura01}. Furthermore, AGB stars are efficient dust
manufacturers, owing to the thermodynamic conditions of their
circumstellar envelope, a favourable environment for dust formation
\citep{fg01, fg02, fg06}. 

The evolution through the AGB is as complex as it is important, owing to the
delicate interface between the thermodynamic structure of the compact,
degenerate core and the tenuous, expanded convective envelope. In the
stellar layers separating these two regions, the main physical
variables, i.e. pressure, temperature and density, drop by several
orders of magnitude, thus rendering the physical and
numerical description of the stellar structure problematic. To this, we add the
uncertain description of convection and mass loss, which is still unknown based
on first principles and thus modelled via empirical prescriptions. These
are the main reasons why, despite the significant progress of AGB
modelling over the last few years (e.g., Karakas \& Lattanzio 2014),
the results are still not completely reliable. The comparison between
theoretical AGB models with observations is crucial in order to
substantially improve the predictive power of the stellar evolution
models.

The Magellanic Clouds (MCs) have been extensively used to test theoretical predictions 
and to derive information on the 
internal structure and on the efficiency of the mechanisms that are altering the surface 
chemical composition of AGB stars. 
The near-- and mid--infrared (IR) colours of the 
stars observed are extremely sensitive to the dust present in the wind; they can thus test 
the description of the dust formation in the circumstellar envelope, which some research
groups have recently coupled with the modelling of the central star \citep{paperI, paperII, 
paperIII, paperIV, nanni13a, nanni13b, nanni14}. This field of research is currently
undergoing significant improvements, with the introduction of chemical--dynamical
models of the circumstellar envelope which are promising replacements of the hydrostatic
approximation currently used \citep{marigo16}.

The comparison between the {\it Spitzer} data and the evolutionary 
sequences allowed the characterisation of the AGB population of the MCs,
in terms of mass, age and metallicity distribution \citep{flavia14, flavia15a, 
flavia15b}. The analysis of the stars exhibiting the largest degree of obscuration in
the MCs was shown to be a valuable indicator not only of the main physical processes
taking place during the AGB evolution, but also of the efficiency with which
dust grains form and grow in the circumstellar envelope \citep{ventura15, ventura16}.

To allow a more exhaustive test of the theoretical models used so far,
it is crucial to extend this kind of analysis to environments with
different metallicities and star formation histories (SFH). This is
now possible, thanks to the mid--IR survey of DUST in Nearby Galaxies
with {\it Spitzer} \citep[DUSTiNGS;][]{boyer15I, boyer15II}, which
provided $3.6\mu$m and $4.5\mu$m {\it Spitzer} imaging of 50 dwarf
galaxies within 1.5 Mpc. The availability of mid--IR data complements
recent HST studies aimed at constraining the lifetimes of AGB stars in
nearby galaxies \citep{girardi10, rosenfield14,
  rosenfield16}. In particular, the combination of near-- and
  mid--IR photometry is a powerful diagnostic of the AGB population of
  dwarf galaxies, with the goal of characterizing the stars observed
  in terms of mass, chemical composition, and progenitor formation
  epoch.  Following the same approach adopted to study the evolved,
  obscured sources of the MCs, we base our analysis on the AGB
  evolutionary sequences that account for dust formation in the wind;
  this is required to correctly interpret the near-- and mid--IR
  fluxes of AGB stars surrounded by dust. We also investigate the most
  obscured stars in the galaxy and predict the overall
  dust--production rate from AGB stars. To date, this is the first
  time that this methodology is applied to galaxies other than Milky
  Way and the MCs, providing a test of the current generation of AGB
  models in different environments.

The paper is organised as follows: the observational sample of AGB
stars considered for the analysis is described in section
\ref{sample}; section \ref{inputs} presents the numerical and physical
inputs used to model the AGB phase and to produce the synthetic
population. The description of the main properties of the AGB phase,
including dust evolution and the impact on the IR emission, is
discussed in section \ref{AGBproperties}; section \ref{interpretation}
present the interpretation of the observations and our conclusions are
offered in section \ref{conclusions}.

\section{IC 1613: infrared observations and SFH}
\label{sample}

In this work, we focus on the irregular dwarf galaxy IC 1613,
 one of the nearest gas--rich dwarf galaxies in the Local Group,
 characterized by low internal reddening and foreground
 contamination. This choice is mainly motivated by the numerous
 population of AGB stars, widely studied in the literature.
 \citet{borissova00} presented the $J$-- and $K$--band photometry of
 IC 1613, studying the distribution of AGB stars in the
 galaxy. \citet{albert00} identified $\sim 200$ carbon stars
 (C--stars) via a narrow--band wide field survey of IC 1613, focused
 on the CN and TiO photometry. \citet{battinelli09}, compared
 narrow--band identification of hundreds of C--stars with their $J-$
 and $K-$ photometry classification, finding a threshold for the
 detection of C--stars. \citet{menzies15} presented a 3yr--survey of
 simultaneous imaging in the $J$, $H$ and $K_s$ of IC 1613, focused
 on the identification of supergiants, oxygen--rich (O--rich) and
 C--stars. Wider and complete surveys of this galaxy in the near--IR
 bands were conducted by \citet{sibbons15} and \citet{chun15}, and
 the former presented a population of $\sim 800$ AGB candidate
 classified on the basis of $JHK$ photometry.  The distance of IC
1613 has been studied by several groups using different methods
\citep{dolphin01,pietrzynski06,tammann11}. We adopt the mean distance
of $\sim760$ kpc, determined by \citet{bernard10} by comparing
 their own measurement using RR Lyrae and Cepheid data to the values
 found in the literature.

We base our analysis of the IC 1613 AGB population on near-- and
mid--IR observations. More specifically, we consider the sample of AGB
candidates observed by \citet{sibbons15} with the Wide Field CAMera
(WFCAM) mounted on UKIRT, in the $J$, $H$ and $K$ bands, covering an
area of 0.8 deg$^{2}$ on the sky, within 4.5 kpc from the galactic
centre. The effects of the internal reddening in the near--IR is
negligible $(E(J - K) = 0.010-0.015$ mag; \citet{sibbons15}),
therefore no correction has been made to account for it. All
magnitudes and colours are corrected for foreground extinction using
the extinction map from \citet{schlegel98}, which gives $E(B - V) =
0.02-0.03$ mag. \citet{sibbons15} flagged each object (as
  stellar, probably stellar, noise like, saturated etc.) in each band,
  on the base of the flux curve-of-growth for a series of apertures
  \citep{sibbons12}. Their criterion to select sources required that a
given object has a magnitude measurement in all the three bands
and that it is classified as stellar or probably stellar in at least
two of the three bands\footnote{This criterion allows to remove the
  majority of the background galaxies, as shown by
  \citet{sibbons15}.}.To exclude the bulk of the foreground objects,
\citet{sibbons15} assumed a color cut of $(J-H)_0>0.64$ mag. This is
not sufficient to remove completely the foreground sources, which are
estimated to contaminate the sample in the central region at the level
of $<1\%$. In order to identify AGB candidates, they considered only stars
brighter than the $K_0=18.28$ mag, i. e. the magnitude of the tip of
the red giant branch (TRGB) at the distance of IC 1613. The resulting catalogue of
AGB candidates is composed of $\sim840$ stars. The AGB sample was
further classified in C-- and O--rich stars, adopting the following
criterion: stars were classified as O--rich if they are confined in
the region $0.75 < (J-K)_0 < 1.15$ mag, while C--stars occupy the
region $(J-K)_0\geq1.15$ mag. Indeed, there is no strict colour
boundary between these two spectral types in $(J - K)$ and the
misclassification of AGB sources in both directions is probable, as we
show later (see section \ref{interpretation}). This classification
represents the red limit for the majority of the O--rich stars better
than the blue limit for the C--type population
\citep{kacharov12,sibbons14}.

We extend the analysis to the mid--IR spectral region in order to
better constrain our AGB models on the basis of the dust
contribution. \citet{boyer15I} present the DUSTiNGS survey of 50 dwarf
galaxies observed in and around the Local Group, which is designed to
detect evolved stars in the dust--producing phase.  IC 1613 was
observed with InfraRed Array Camera in the [3.6] and [4.5] bands
during two epochs, reducing the effect of variability. The level of
extinction in these bands results in a change in magnitude that is
significantly less than the photometric uncertainties, so we have not
corrected for it. Having only the measurement of the [3.6] and [4.5]
magnitude available, it was not possible to reduce the contamination
from background and foreground sources. 
\citet{boyer15I} adopted TRGB magnitude $M_{3.6}=-6$~mag for all the
50 galaxies, considering only star brighter than this limit as AGB
stars. While they could not classify all detected AGB stars with only
the mid-IR data, they did separate a subset of individual AGB stars
from foreground and background sources by their magnitude changes
between two epochs.

We cross--correlated the sample of AGB candidates by \citet{sibbons15}
with stars from the DUSTiNGS ``Good Source"
catalogue \footnote{The``Good Source" catalogue is culled to include
  only high--confidence point sources and reliable measurements} by
\citet{boyer15I}, using a matching radius of 1.2 arcsec. In this way
we were able to select 755 objects, probably AGB stars, observed in
both near-- and mid--IR bands. 10\% of the AGB candidates observed by
\citet{sibbons15} are not matched with any star in the Boyer's
sample. The reason is that the field of view observed by UKIRT is
wider than the {\it Spitzer} one: therefore, the majority (90\%) of
the stars observed by UKIRT that are not in the {\it Spitzer} sample
fall in the region not covered by the DUSTiNGS field of view. Note
that the most obscured stars are excluded by the criterion assumed by
\citet{sibbons15} to select AGB candidates because they are not
detected in the near--IR bands. However, these stars are present in
the DUSTiNGS catalogue (see Section \ref{xAGB}).

Recently, \citet{skillman14} presented a study of IC 1613 based on
HST/ACS observations. In an area that covers 9\% of the field, which
is considered representative of the entire galaxy, they find an early
phase of star formation, lasting 5--6Gyr, characterized by an
approximately constant SFR ($\sim 10^{-9}$
$M_{\odot}yr^{-1}pc^{-2}$). This is followed by an epoch, started
$\sim8$Gyr ago, during which the SFR has been a factor $\sim2$
smaller.  The recent era is characterized by a moderate and constant
SFR ($\sim0.4 \times 10^{-9}$ $M_{\odot}yr^{-1}pc^{-2}$). Concerning
the age--metallicity relation (AMR), \citet{skillman14} derive a
constant increase of the metallicity ($Z$) during the evolution of the
galaxy; the values of $Z$ span the range $4\times10^{-4} \leq Z\leq 4
\times 10^{-3}$. In this work we use the SFR and AMR by
\citet{skillman14} to construct the synthetic population of AGB stars,
as described in section \ref{interpretation}.

\section{Numerical and physical inputs}
\label{inputs}
\subsection{Stellar evolution models}
This work is based on evolutionary sequences calculated with the ATON
code, extensively described in \citet{ventura98}; the interested
reader may find in \citet{ventura09} a list of the most recent
chemical and physical input. The models trace the pre--main sequence
to the almost complete loss of the external mantle.  The metallicities
used are $Z=10^{-3}$, $Z=2\times 10^{-3}$ and $Z=4\times 10^{-3}$.
The mixtures adopted are taken from \citet{gs98}; for the lowest
metallicity cases, we use $[\alpha/Fe]=+0.4$ and for $Z=4\times
10^{-3}$, we adopted $[\alpha/Fe]=+0.2$.

We adopt the physical input that is most relevant for this work, described as follows: 

\begin{itemize} 

\item{The convective instability is described by means of the Full
Spectrum of Turbulence (FST) model developed by \citet{cm91}. This prescription was shown to favour strong
hot bottom burning (HBB) conditions during the AGB phase of  $M > 3~\Msun$ stars \citep{vd05}. Other models in the literature, based on the traditional mixing length theory (MLT), find that the lower limit in mass to reach HBB condition is $\sim5M_{\odot}$ \citep{karakas14, marigo13}}

\item{Nuclear burning and mixing of chemicals are self--consistently coupled,
in a diffusive--like scheme. Overshoot of convective eddies into radiatively stable regions
is described by means of an exponential decay of velocities from the convective/radiative
interface, fixed by the Schwarzschild criterion. The e--folding distance is assumed to
be $0.002H_p$ (where $H_p$ is the pressure scale height calculated at the formal boundary
of convection), in agreement with the calibration based on the observed luminosity
function of C--stars in the Large Magellanic Cloud (LMC), given by \citet{paperIV}. 
}

\item{The mass--loss rate for oxygen--rich models is determined via the 
\citet{blocker95} treatment; we set the free parameter $\eta_{R}$ to $0.02$, according to the calibration based on the luminosity function of lithium-rich stars in the MCs, given in \citet{ventura00}. For C--stars we use the results from the Berlin group
\citep{wachter02, wachter08}. To account for the significant mass loss suffered by $M<1M_{\odot}$ stars during the pre-AGB phases, we determined the mass at the beginning of the AGB evolution for these objects adopting the prescription by \citet{rosenfield14}.}

\item{The molecular opacities in the low--temperature regime (below 
$10^4$ K) are calculated by means of the AESOPUS tool \citep{marigo09}. 
The opacities are suitably constructed to follow the changes in the chemical composition 
of the envelope. This was shown to significantly affect the evolution following the
achievement of the C--star stage \citep{vm09, vm10}.} 

\end{itemize}

\subsection{Dust formation in the winds of AGB stars}
\label{dustmodel}
The growth of dust particles in the circumstellar envelope is modelled according to
the scheme introduced by the Heidelberg group \citep{gs85, gs99, 
fg01, fg02, fg06, zhukovska08}. The overall set of equations used and the physical 
assumptions behind this model are extensively described in the afore mentioned papers and
by the previous works by our group \citep{paperI, paperII, paperIII, flavia14b,
paperIV}.  

The wind is assumed to expand isotropically, with constant velocity,
from the surface of the star; once the gas particles enter the
condensation zone, the wind is accelerated owing to the effects of
radiation pressure on the newly formed grains.  In this model, the
dynamical conditions of the wind are determined on the basis of the
mass and momentum conservation, whereas the growth of dust particles
is found on the basis of the gas density and the thermal velocity in
the inner border of the condensation region. Gas dynamics and dust
grain growth are coupled via the extinction coefficient in the
equation of momentum conservation, which depends on the number density
and size of the grains formed.

The dust species taken into account depend on the surface chemical
composition of the star.  In carbon rich environments
($C/O > 1$), formation of solid carbon, silicon carbide (SiC) and solid
iron are considered, whereas in oxygen--rich atmospheres we account
for the formation of silicates, alumina dust and solid iron.

For any stage during the AGB evolution, the dust formed in the wind is determined
self--consistently based on the physical parameters of the star (luminosity, mass,
effective temperature and mass--loss rate) and on the surface chemical composition.

\subsection{Synthetic spectra}
\label{spectramodel}
The near-- and mid--IR magnitudes used in the present analysis are
found by a two--step process. First, on the basis of the values of
mass, luminosity, effective temperature, mass--loss rate and surface
chemical composition of a given AGB phase, the dust-formation model
described in Section \ref{dustmodel} is applied to determine the size
of the dust grains formed and the optical depth (here we use the value
at $10~\mu$m, $\tau_{10}$).  These ingredients are used by the code
DUSTY \citep{dusty} to calculate the synthetic spectra of each
selected point along the evolutionary sequence. The input
  radiation from the central star was obtained by interpolating in
  gravity and effective temperature among the appropriate tables of
  the same metallicity: we used the NEXTGEN atmospheres
  \citep{hauschildt99} for oxygen--rich stars and the COMARCS
  atmospheres \citep{grams} for carbon stars. In the latter case we
  interpolated among the C/O values. The magnitudes in the various
bands are obtained by convolution with the appropriate transmission
curves.  DUSTY needs as input parameters the effective temperature of
the star, the radial profile of the gas density and the dust
composition of the wind, in terms of the percentage of the various
species present and of the size of the dust particles formed. All
these quantities are known based on the results of stellar evolution
and of the description of the wind. The interested reader may find a
detailed description of this process in \citet{flavia15a} (Section
2.3).

\begin{figure*}
\begin{minipage}{0.33\textwidth}
\resizebox{1.\hsize}{!}{\includegraphics{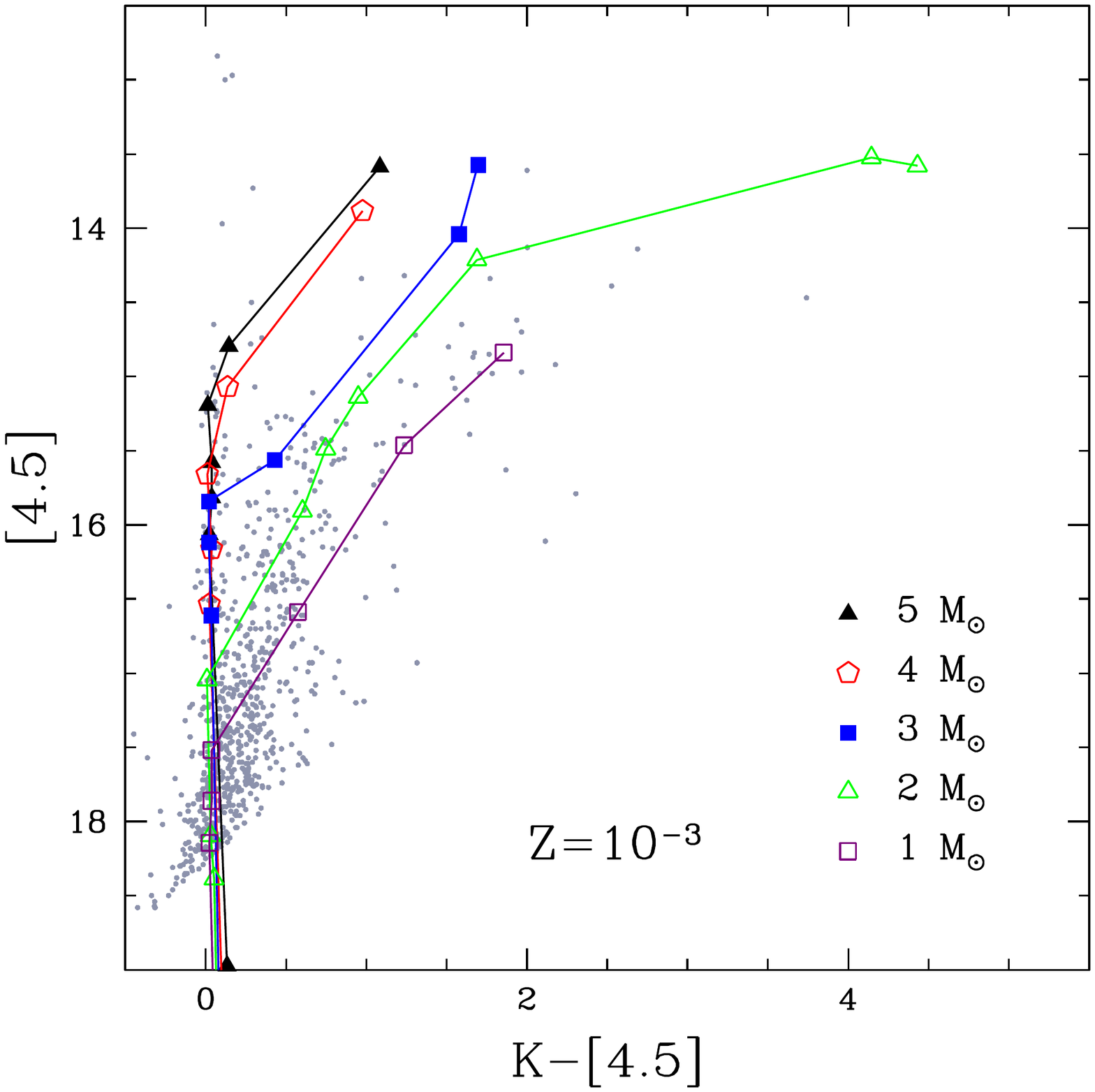}}
\end{minipage}
\begin{minipage}{0.33\textwidth}
\resizebox{1.\hsize}{!}{\includegraphics{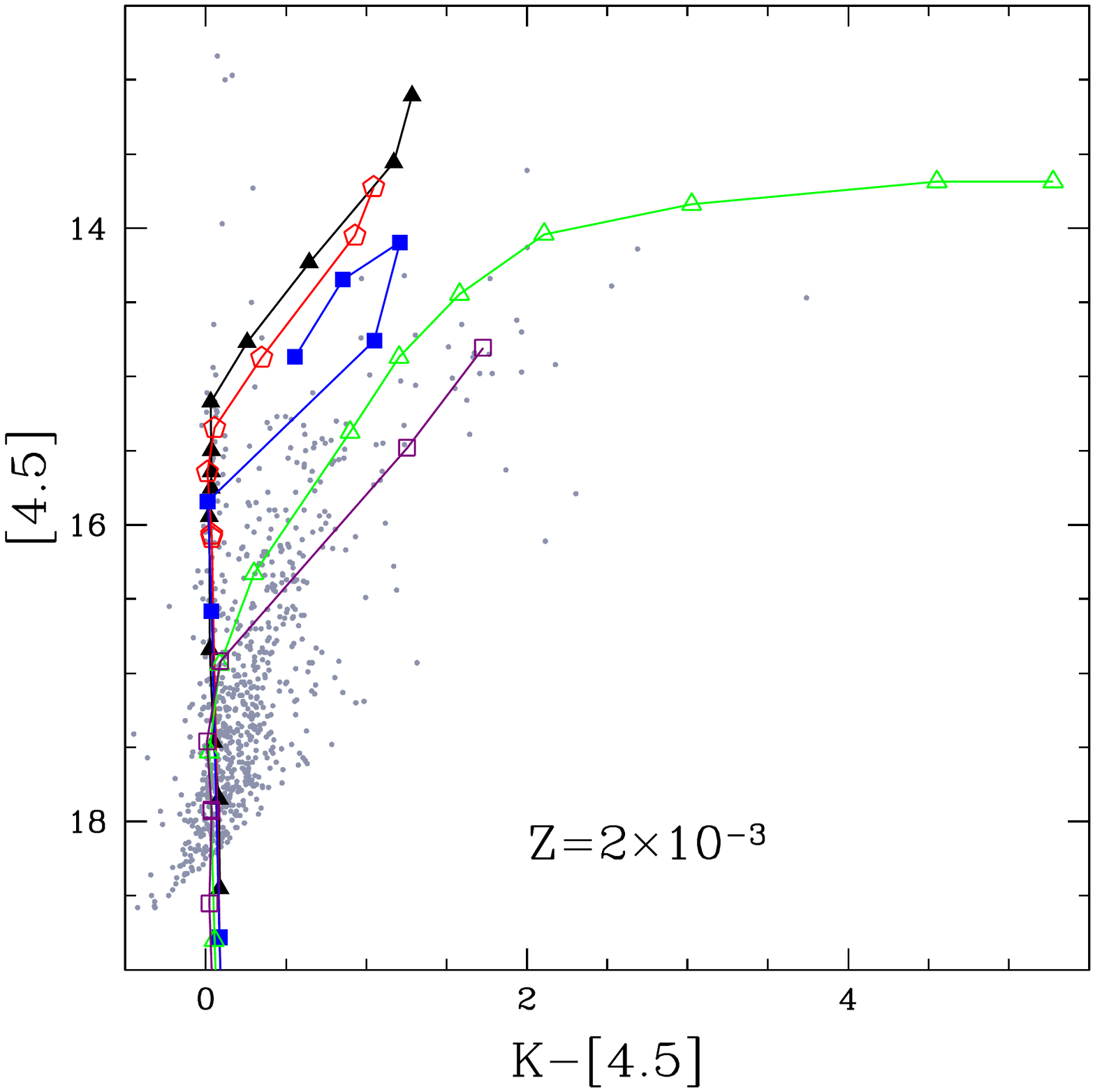}}
\end{minipage}
\begin{minipage}{0.33\textwidth}
\resizebox{1.\hsize}{!}{\includegraphics{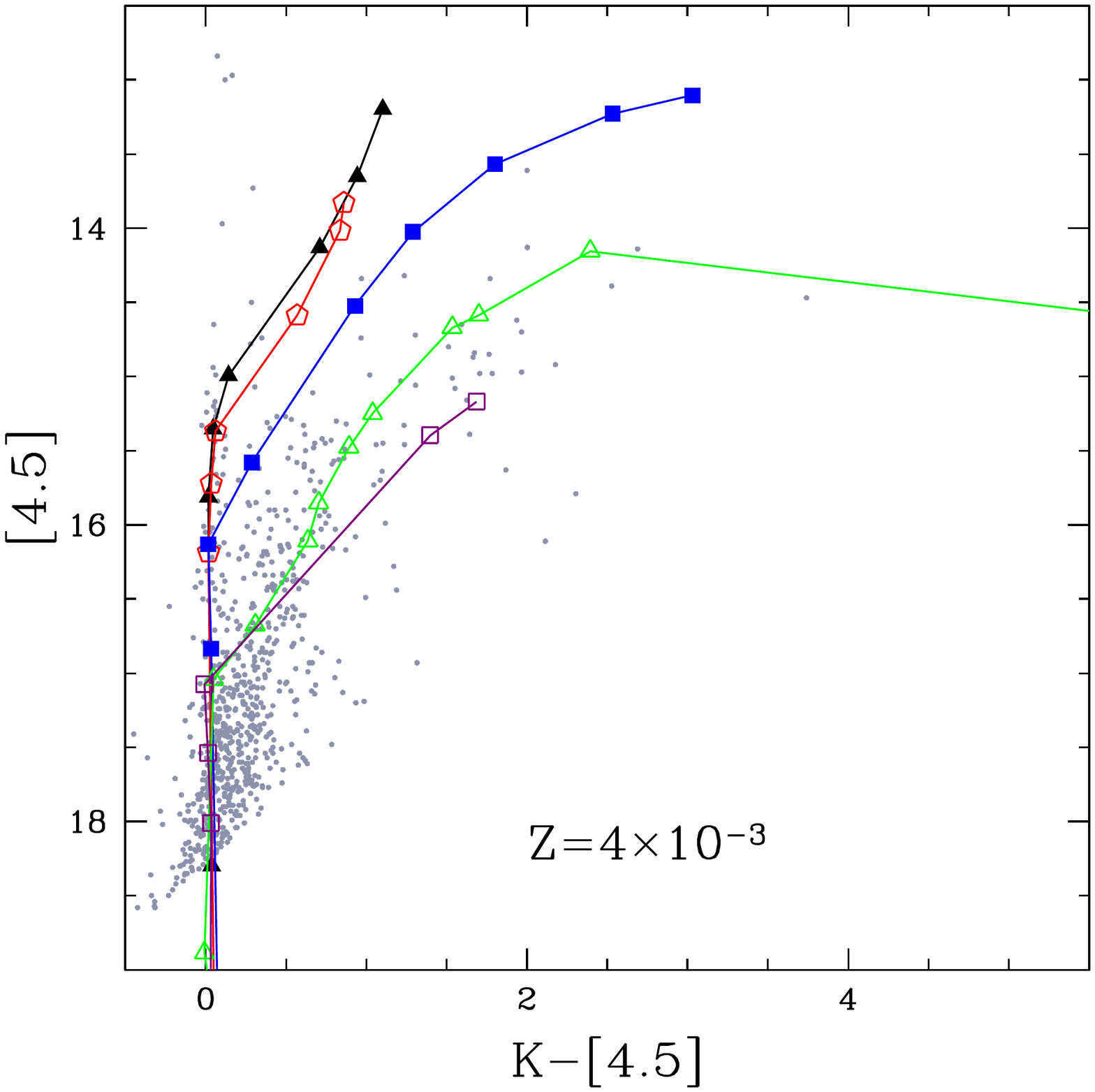}}
\end{minipage}
\vskip-30pt
\caption{Evolutionary tracks in the colour--magnitude ($K - [4.5],
  [4.5]$) diagram of AGB stars of different initial mass (purple open
  square: 1$M_{\odot}$; green open triangle: 2$M_{\odot}$; blue full
  square: 3$M_{\odot}$; red open circle: 4$M_{\odot}$; black full
  triangle: 5$M_{\odot}$) and metallicity $Z=10^{-3}$ (left panel),
  $Z=2\times 10^{-3}$ (middle) and $Z=4\times 10^{-3}$ (right). The
  grey points represent observational data of the AGB candidate sample
  in IC 1613 from \citet{sibbons15} and \citet{boyer15I}, as described
  in Section \ref{sample}. Note that the $K$ magnitude is corrected
  for foreground extinction in all figures. }
\label{CMD_tracks}
\end{figure*}

\section{The evolution through the AGB phase}
\label{AGBproperties}
The $Z=1.4\times 10^{-3}$ models used here are extensively discussed
in \citet{ventura13,paperIV,ventura14b}. We direct the interested
reader to these papers for a full presentation of the evolutionary
sequences. The $Z=2\times 10^{-3}$ models have been computed
specifically for the present investigation. We consider the
  entire AGB evolution, which begins with the ignition of shell helium
  burning (early--AGB phase) and extends to the thermal pulse (TP) AGB phase.

\subsection{The role of the initial mass}
\label{evolution}
The physical and chemical evolution through the AGB phase is primarily affected
by the initial mass of the star, as the latter quantity determines the core mass 
at the beginning of the TP phase, which is the key quantity in 
driving the AGB evolution \citep{herwig05, karakas14}.

Models with mass above $\sim 3~M_{\odot}$ experience HBB,
consisting of an advanced proton capture nucleosynthesis at the bottom of the convective 
envelope \citep{renzini81}. This nuclear activity requires that the temperature at the
base of the external mantle ($T_{bce}$) exceeds $\sim 30$ MK. The activation of HBB favours a 
significant increase in the stellar luminosity \citep{blocker91}, which makes the core 
mass vs. luminosity relationship much steeper than that predicted by the classic 
\citet{paczinski} law. HBB affects the overall duration of the TP--AGB phase, because the higher 
luminosity favours a faster loss of the external mantle 
\citep{vd05}. On the chemical side, the ignition of HBB triggers the activation of CN 
cycling, with the synthesis of nitrogen at the expense of carbon; this process requires
temperatures $T_{bce} \geq 40$ MK. Stars of mass above $\sim 5~M_{\odot}$ evolve at
$T_{bce} > 80$ MK, thus experiencing an even more advanced nucleosynthesis, with the
activation of the full CNO cycle. The surface chemical composition of 
these stars will be carbon and oxygen poor, and enriched in nitrogen. The achievement of 
the C--star stage in the stars undergoing HBB is prevented by the proton capture 
nucleosynthesis at which carbon nuclei are exposed in the external envelope. 

Stars of initial mass below $3~M_{\odot}$ do not experience any HBB;
their surface chemical composition is altered only by the third dredge
up (TDU) phenomenon \citep{iben83}.  The TDU is characterised by the
inward penetration of the base of the convective envelope following
each TP; the external mantle may thus reach internal layers undergoing $3\alpha$ nucleosynthesis and greatly enriched in carbon. Oxygen is also
expected to be enhanced in this region of the star, though at a
smaller extent than carbon. Owing to repeated TDU events, which
eventually lead to the $C/O>1$ condition, the stars with mass in the
range $1~\Msun \leq M \leq 3~\Msun$ become C--stars.  The upper limit
of this mass interval is determined by the achievement of HBB
conditions, whereas stars of mass below $1~\Msun$ loose the external
mantle before becoming C--stars.  The surface carbon enrichment
increases with the mass of the star, because models of higher mass
experience more TPs (thus, more TDU events) before the envelope is
entirely lost. Note that the upper limit in mass for the
  achievement of the C--star stage is dependent on the convection
  modelling. The $\sim3M_{\odot}$value given above is found when the
  FST description is adopted. Other investigators, using the MLT
  approach \citep{karakas14,marigo13} find a wider range in mass for
  carbon stars. Unfortunately, the present study cannot be used to
  discriminate among the different descriptions, because the relative
  number of $\sim5M_{\odot}$ C-stars expected (see Section
  \ref{interpretation} and Fig. \ref{isto_mass}), if any, is
  significantly smaller than lower-mass counterparts to allow any
  statistics of the distribution of stars along the C-star sequence.

\subsection{The effects of metallicity}
The metallicity of the stars has a significant impact on the results obtained; this is 
because the efficiency of the mechanisms potentially able to alter the surface chemical
composition, namely TDU and HBB, changes with the metallicity. An additional reason is
that the core mass vs. initial mass relationship becomes steeper when the metallicity
decreases\footnote{In the low-mass ($\leq2\Msun$) domain, the differences are due to the higher efficiency of the H-burning shell in lower metallicity stars, when evolving along the RGB; this makes the core to grow faster, thus to higher core mass in the following phases. For $M>3\Msun$ models, the core masses of lower metallicity stars are larger, because they experience a more penetrating second dredge-up.}. 

On the physical side, because lower metallicity stars evolve on more
massive cores, the upper limit (in mass) of the stars experiencing the
AGB evolution and thus avoiding core collapse, changes from $7.5~\Msun$
for $Z=10^{-3}$ stars to $8~\Msun$ for $Z=4\times 10^{-3}$.  The
lower limit for the ignition of HBB also increases with Z: it changes
from $2.5~\Msun$ in the $Z=10^{-3}$ case to $3~\Msun$ for
$Z=4\times 10^{-3}$.

The surface chemistry of the stars experiencing HBB is significantly affected by the
metallicity because lower--Z stars are exposed to a more advanced nucleosynthesis at the
base of the envelope. This particularly affects the oxygen content, which is 
destroyed more easily in lower metallicity models \citep[e.g., see Fig.~3 in][]{ventura13}.

In the low--mass star domain the main effect of metallicity is on the lowest mass of the stars
eventually reaching the C--star stage: in the $Z=10^{-3}$ case all stars of mass above
$1~\Msun$ become C--stars, whereas for $Z=4\times 10^{-3}$ this lower limit is
$1.25~\Msun$. The reason for this trend with metallicity is twofold: a) stars of lower Z
achieve more easily the C--star stage, because the initial oxygen is smaller; b) the
efficiency of TDU is higher in models of lower metallicity \citep{boothroyd88}.

\subsection{Dust production in AGB stars}
The winds of AGB stars prove an extremely favourable environment to dust formation,
because they are extremely cool and sufficiently dense to allow condensation of a large 
number of gas molecules into dust \citep{fg01, fg02, fg06}.

The dust production by the AGB models used in the present analysis are extensively 
discussed in \citet{paperI, paperII}, \citet{flavia14b}, and \citet{paperIII}. An overall view of the results for
various masses and metallicities is presented in \citet{paperIV}. Here we provide a
short summary of the main findings of those works, most relevant for the present
investigation.

The threshold mass of $\sim 3~\Msun$ for the activation of HBB also plays an important 
role in the type of dust produced: this is because the dust formed depends on the atmospheric C/O 
ratio, with carbon-rich stars producing carbonaceous dust ($M<3~\Msun$) and oxygen--rich AGB stars ($M > 3~\Msun$) forming silicates\footnote{Note that we derived self-consistently the relative percentages of iron-poor/iron-rich silicates formed \citep[see eq. 8 and 9 in][]{paperIV}. In all the cases of interest here we find that silicates produced are iron-poor.} and alumina dust \citep{paperI}.

In the latter case, the amount of silicates and alumina dust formed 
is approximately
proportional to the metallicity, because the abundance of silicon and aluminium scales 
almost linearly with Z. The same holds for the SiC formed in the lower mass counterparts, 
which is also dependent on the silicon available in the envelope. Conversely, the solid 
carbon formed is at first approximation independent of Z, because the carbon transported to the 
surface via TDU is formed in situ.

In the higher mass domain, the quantity of dust formed increases with the initial mass
of the star, because the density of the wind generally increases with mass, at least 
for the stars experiencing HBB. In lower mass stars the amount of carbon dust
formed increases with the initial mass, as the carbon accumulated to the surface is
larger in higher mass models \citep[see Fig.~9 and ~10 in][]{paperIV}.

\subsection{The infrared properties of AGB stars}
\label{tracks}
The degree of obscuration of AGB stars depends on the type and
quantity of dust formed in the circumstellar envelope. Dust particles
reprocess the radiation emitted by the central star to the IR, thus
shifting the peak of the spectral energy distribution to longer
wavelengths.

In stars experiencing HBB, the degree of obscuration increases during
the initial part of the TP--AGB evolution, because the growth of the
core mass leads to stronger HBB conditions, higher mass--loss rates,
and higher density winds, resulting in higher rates of silicate and
alumina dust grain growth \citep[see Fig.~2 in][]{flavia15b}.  The
largest optical depths (of order unity) are reached in conjunction
with the phase when HBB is strongest, before the gradual loss of the
envelope provokes a general cooling of the whole external region of
the star. The fraction of the TP--AGB evolution during which a star is
obscured depends on the mass and metallicity.  Massive AGB stars, with
mass above $\sim 5-6~\Msun$, experience strong HBB at the first TPs,
thus evolving with a large degree of obscuration for the majority of
the TP--AGB phase. Conversely, stars of smaller mass ($3\Msun<M<5\Msun$)
experience an initial TP--AGB evolution with scarce amount of dust in
their wind; the fraction of TP--AGB life during which they are
obscured is below $\sim 50\%$. The metallicity trend is
therefore straightforward: because higher--Z, massive TP--AGB stars produce more
dust, the fraction of the TP--AGB life when the stars are obscured is
larger in higher--Z objects.

Stars of mass below $3~\Msun$ evolve initially with no dust, as
formation of silicates in the phases previous to the achievement of
the C--star phase is negligible.  $\tau_{10}$ increases significantly
at the beginning of the C--star phase, and becomes larger as more
carbon is accumulated to the surface. This phase is longer at lower
metallicity, where the $C/O>1$ condition is reached
more easily \citep[see Fig.~2 in][]{flavia15b}. Because more massive
stars accumulate more carbon, the stars of initial mass around $\sim
2.5~\Msun$ are those reaching the largest degree of obscuration
($\tau_{10} \sim 3$).

Fig. \ref{CMD_tracks} shows the path traced by the evolutionary tracks of AGB models of various
mass and metallicity in the color--magnitude diagram (CMD; $K-[4.5],[4.5]$). The three panels 
refer to the metallicities $Z=1,2,4 \times 10^{-3}$ and show the tracks of low--mass
stars experiencing TDU and higher mass objects, whose external mantle is exposed to HBB. 
All the tracks move vertically in the initial early- and TP--AGB phases, because little dust forms and 
the degree of obscuration is small. The luminosity of the stars, hence the $[4.5]$ flux,
increases as the core mass grows. 

The tracks of stars of initial mass below 
$\sim 1~\Msun$ keep vertical for the whole AGB phase, because they never become carbon
stars, and the amount of silicates produced during their evolution is too small to allow
a significant degree of obscuration.

The stars of mass in the range 1--3~$\Msun$, as discussed previously, 
evolve initially as oxygen--rich objects, then become C--stars once the C/O ratio 
exceeds unity. After the C--star stage is reached,
carbon dust forms and the tracks moves rightwards in the CMD. The $K-[4.5]$ colour
becomes redder as the stars evolve through the TP--AGB, due to the progressive
carbon enrichment of the surface layers, in turn favoured by repeated TDU events. This is the case for $\sim 2-2.5M_{\odot}$ stars, which accumulate the largest amount of carbon in the external regions, favoring the highest dust production. At a first approximation this mechanism occurs independently from $Z$.
The increase in the $4.5 \mu$m flux is
associated to the progressive shift of the SED to longer wavelengths, while the
overall luminosity of the star keeps approximately constant in the latest TP--AGB phases. 
We stress here that the evolutionary times become progressively shorter as the degree of
obscuration (hence, the $K-[4.5]$ colour) increases because the presence of carbonaceous 
dust, owing to the effects of radiation pressure acting on dust grains, favours the 
increase in the mass-loss rate.

For stars of initial mass above $3~M_{\odot}$, the mass-loss rate in
the initial TP--AGB phase is not sufficient to form the amount of dust
necessary to provoke a significant degree of obscuration, thus the
tracks move vertically in the CMD.  In more advanced phases, the
tracks move rightwards as dust forms. Unlike their lower mass
counterparts, in this case the $K-[4.5]$ colours reach a maximum
during the phase of strongest HBB and decreases
afterwards\footnote{For clarity, in Fig.\ref{CMD_tracks} we shown only
  the evolution until the maximum $K-[4.5]$ colours reached by the
  star}.

\begin{figure*}
\begin{minipage}{0.49\textwidth}
\resizebox{1.\hsize}{!}{\includegraphics{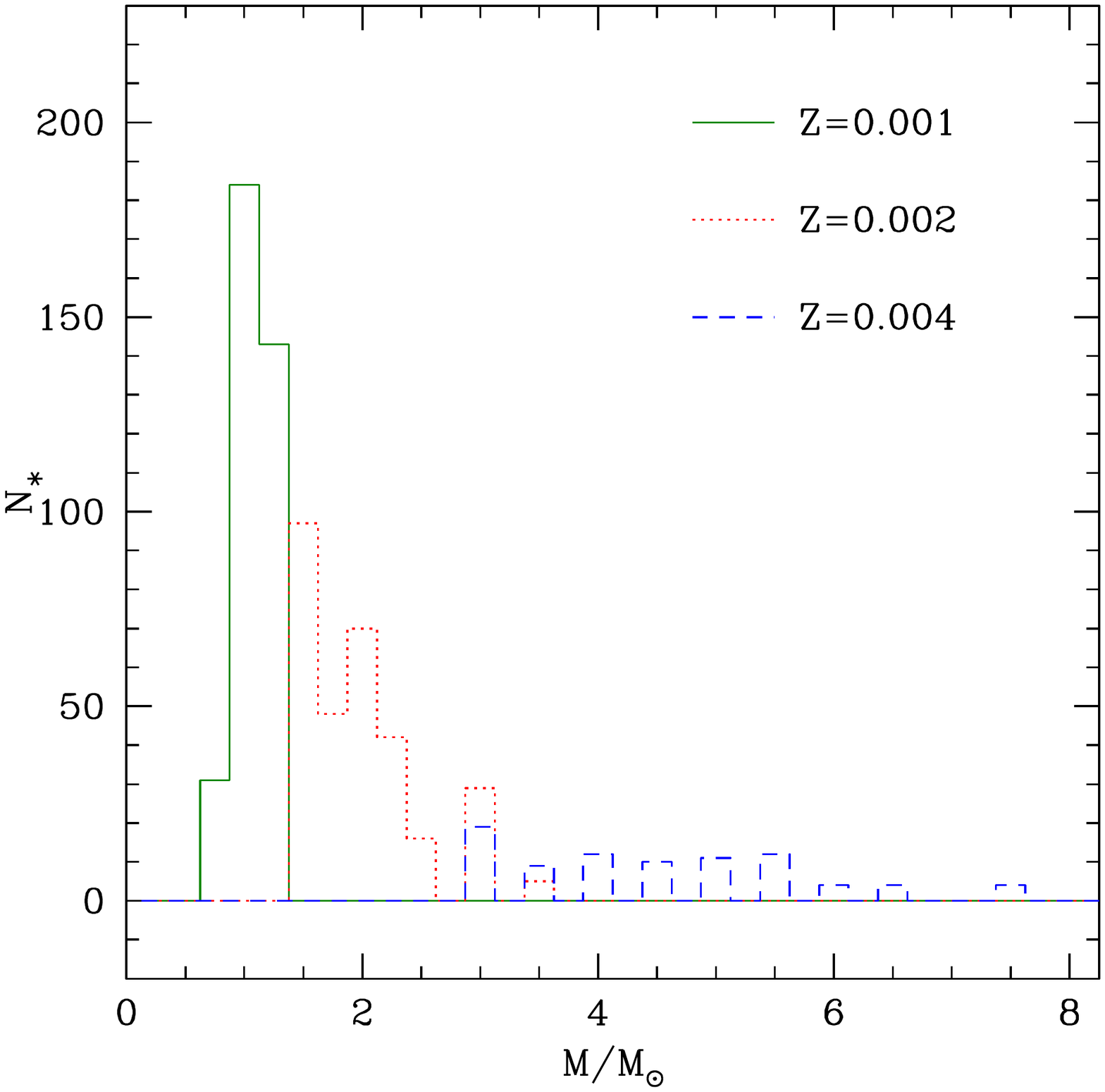}}
\end{minipage}
\begin{minipage}{0.49\textwidth}
\resizebox{1.\hsize}{!}{\includegraphics{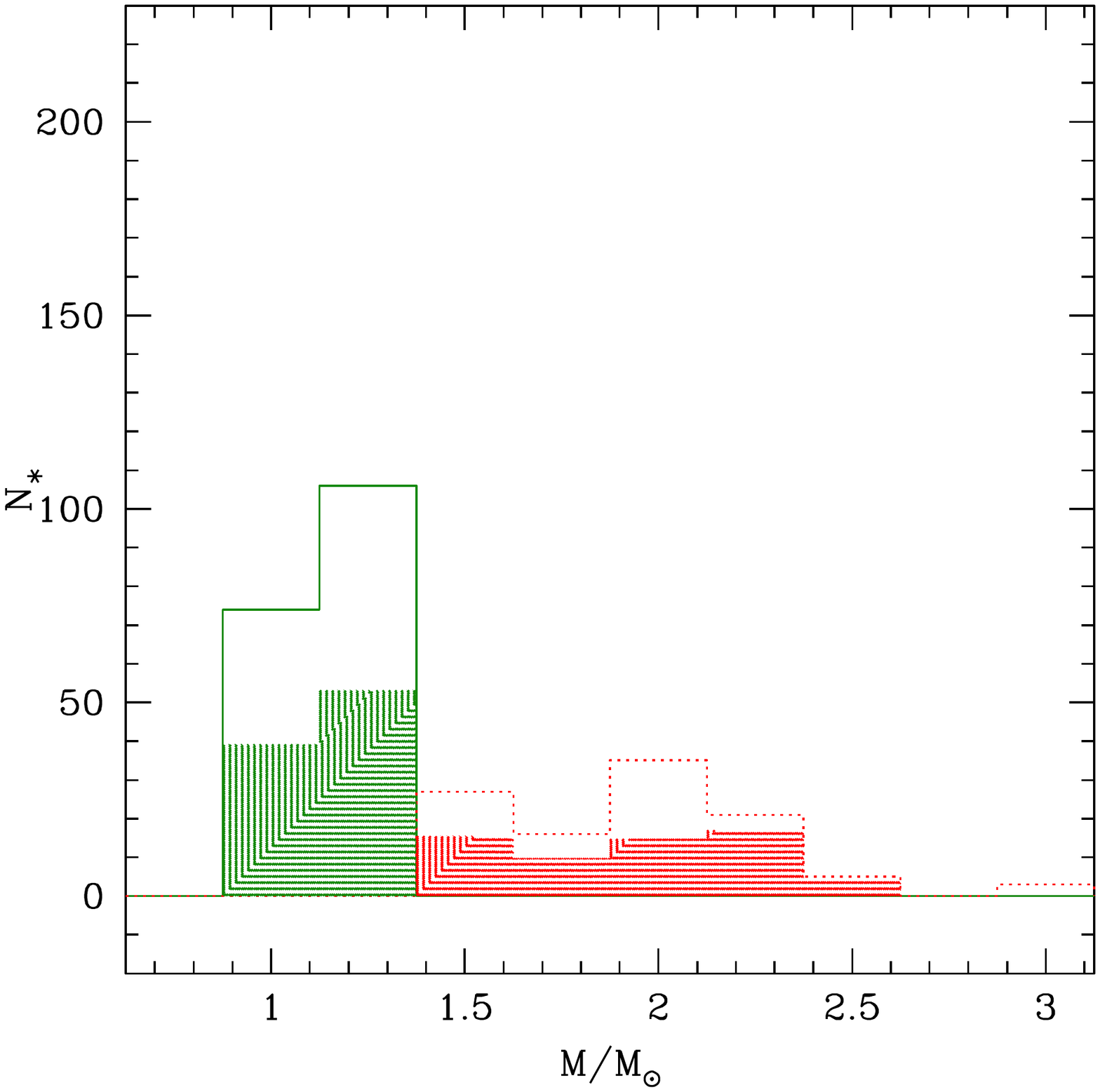}}
\end{minipage}
\vskip-50pt
\caption{Left panel: the mass distribution of the AGB population of IC
  1613, according to our synthetic modelling. The various masses are
  separated among the three different metallicities used in the
  present analysis: $Z=10^{-3}$ (solid green), $Z=2\times 10^{-3}$
  (dotted red) and $Z=4\times 10^{-3}$ (dashed blue). Right panel: the
  mass distribution of C--stars from the synthetic AGB population,
  with the same color code used in the left panel. The filled
  histogram refers to C--stars with $K-[4.5]>0.5$~mag.}
\label{isto_mass}
\end{figure*}

\section{The AGB population in IC 1613}
\label{interpretation}

\begin{figure*}
\begin{minipage}{0.49\textwidth}
\resizebox{1.\hsize}{!}{\includegraphics{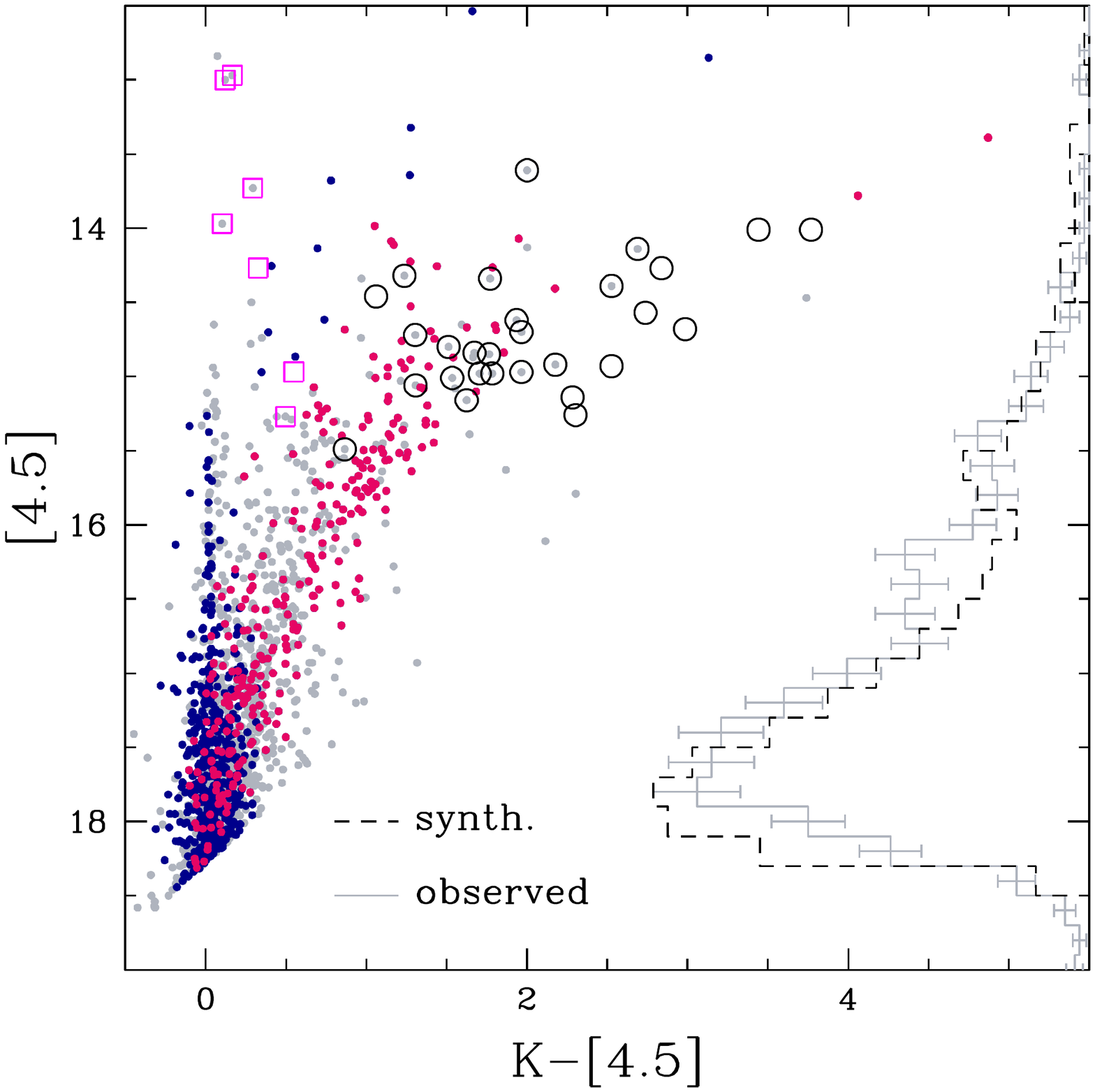}}
\end{minipage}
\begin{minipage}{0.49\textwidth}
\resizebox{1.\hsize}{!}{\includegraphics{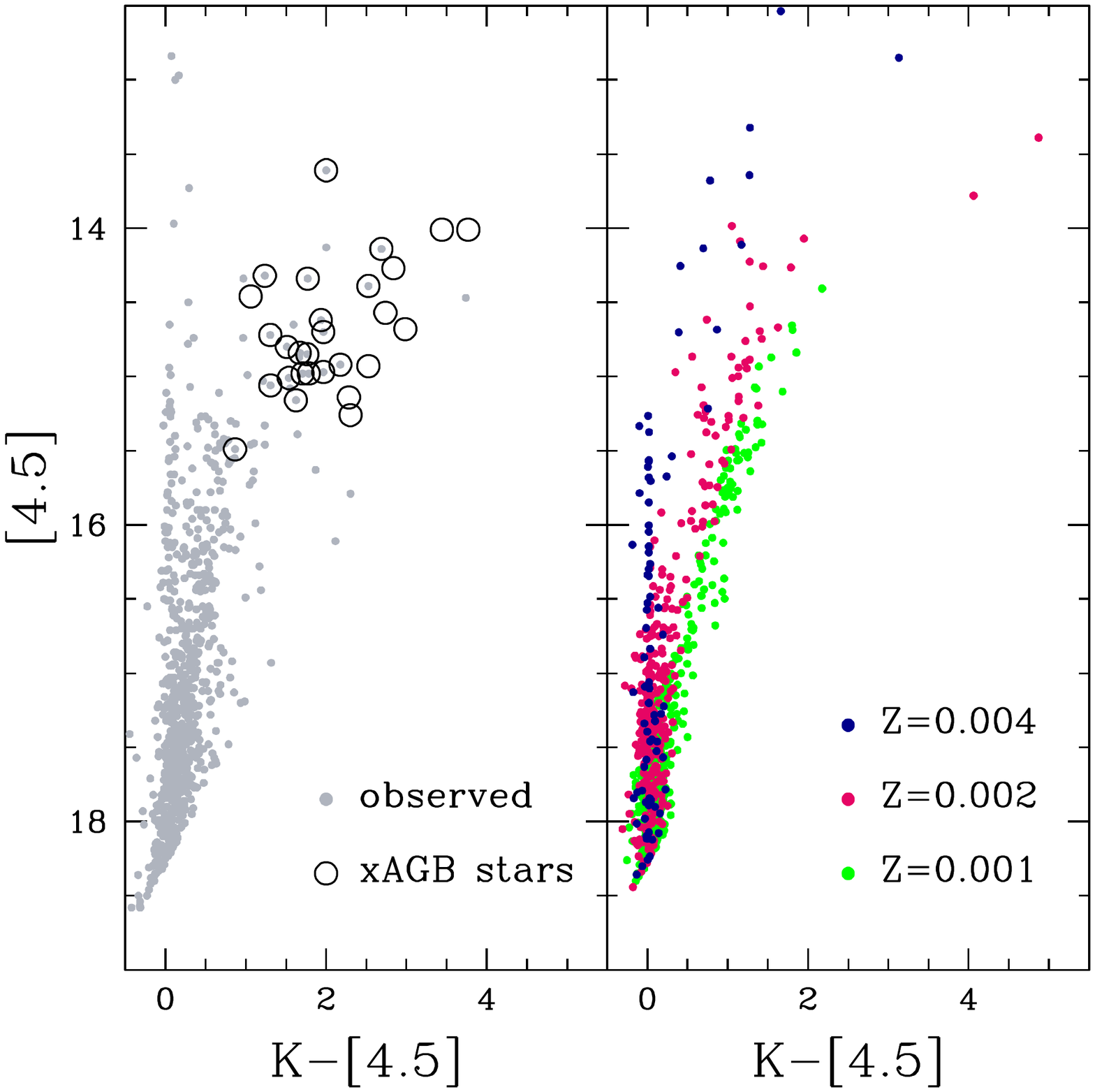}}
\end{minipage}
\vskip-50pt
\caption{Left panel: the colour--magnitude ($K - [4.5], [4.5]$)
  diagram of the AGB candidate sample (grey points); the blue and red
  dots indicate respectively the O--rich and C--stars from synthetic
  modelling. The histograms on the y--axis represent the distribution
  of observed (solid grey) and synthetic (dashed black) stars in [4.5]
  magnitude. RSG stars (open squares) classified by \citet{menzies15}
  and xAGB stars (open circle) identified by \citet{boyer15II} are
  also shown. Right panel: the colour--magnitude ($K - [4.5], [4.5]$)
  diagram of the AGB candidate sample (grey) and xAGB stars (open
  cirlces) on the left; on the right, the synthetic population divided
  in the three metallicities considered: $Z=10^{-3}$ (green),
  $Z=2\times 10^{-3}$ (red) and $Z=4\times 10^{-3}$ (blue).}
\label{CMD}
\end{figure*}

To characterise the AGB population of IC 1613, we used the
evolutionary sequences described in Section \ref{tracks} to produce a
synthetic distribution of the AGB population. Our analysis is based
essentially on the $(K-[4.5],[4.5])$ colour--magnitude diagram; this
choice is motivated by the clear separation of O-- and C--rich
stars in this plane. Additionally, reliable $K$
magnitudes are available even for the stars with a large degree of
obscuration, thus rendering use of the $K$ fluxes more reliable than
the $J$ band. Following the procedure described in \citet{flavia15a,
  flavia15b}, the number of stars extracted for each mass and
metallicity depends on the SFR \citep[from][]{skillman14} and the
initial mass function (IMF, here we used a Salpeter law, with
$x=-1.3$) at the epoch of star formation and on the overall
duration of the AGB phase.We consider the photometric uncertainties
  \citep{sibbons15,boyer15I} in the calculation of the predicted AGB
  colors and magnitudes.

Concerning the metallicity distribution, based on the results by
\citet{skillman14}, we assumed that the stars older than 4 Gyr formed
with $Z=10^{-3}$. Similarly, we use $Z=2\times 10^{-3}$
for ages in the range 0.5--4 Gyrs and $Z=4\times 10^{-3}$ in the most
recent epochs\footnote{According to \citet{skillman14}, the
  metallicity of stars older than $\sim 8$Gyr should be smaller than
  $Z=10^{-3}$, which would require use of $\sim 1M_{\odot}$ models of
  $Z<10^{-3}$. However, as shown by \citet{paperIII}, the main
  evolution and dust production properties of low--mass,
  low--metallicity objects is not significantly sensitive on
  $Z$. Therefore, we may safely assume a minimum metallicity of
  $Z=10^{-3}$ in the present analysis.}.

The left panel of Fig. \ref{isto_mass} shows the distribution of the
stars extracted, as a function of the initial mass and
metallicity. For consistency with the analysis by \citet{sibbons15},
only the synthetic stars with $K<18.28$ mag are shown. Approximately
half of the stars belong to the $Z=10^{-3}$ component, and their mass
distribution peaks around $1M_{\odot}$, corresponding to stars formed
$\sim5.5$ Gyr ago. A significant fraction ($\sim40\%$) is from stars
with $Z=2\times 10^{-3}$ and mass in the range
$1.5-2M_{\odot}$, formed between 800 Myr and 1.8 Gyr ago. The
theoretical population is completed ($\sim10\%$) by higher metallicity
($Z=4\times 10^{-3}$) objects, with masses uniformly distributed in
the range 3--5.5$M_{\odot}$, formed between 70 and 300 Myrs ago. We also predict few
objects of mass $6 \leq M/M_{\odot} \leq 7.5$,
accounting for a negligible fraction of the total population.

\begin{figure*}
\begin{minipage}{0.33\textwidth}
\resizebox{1.\hsize}{!}{\includegraphics{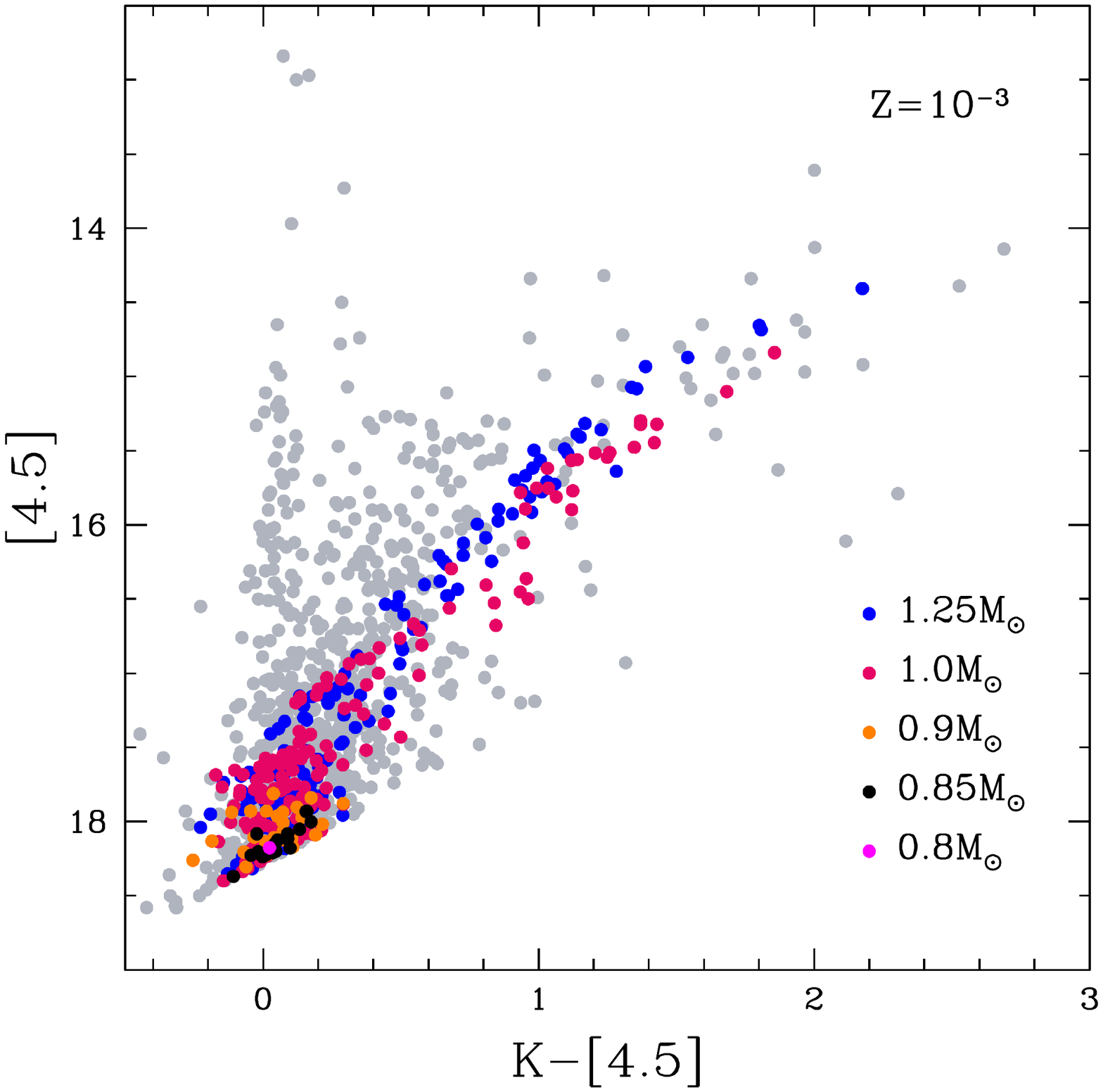}}
\end{minipage}
\begin{minipage}{0.33\textwidth}
\resizebox{1.\hsize}{!}{\includegraphics{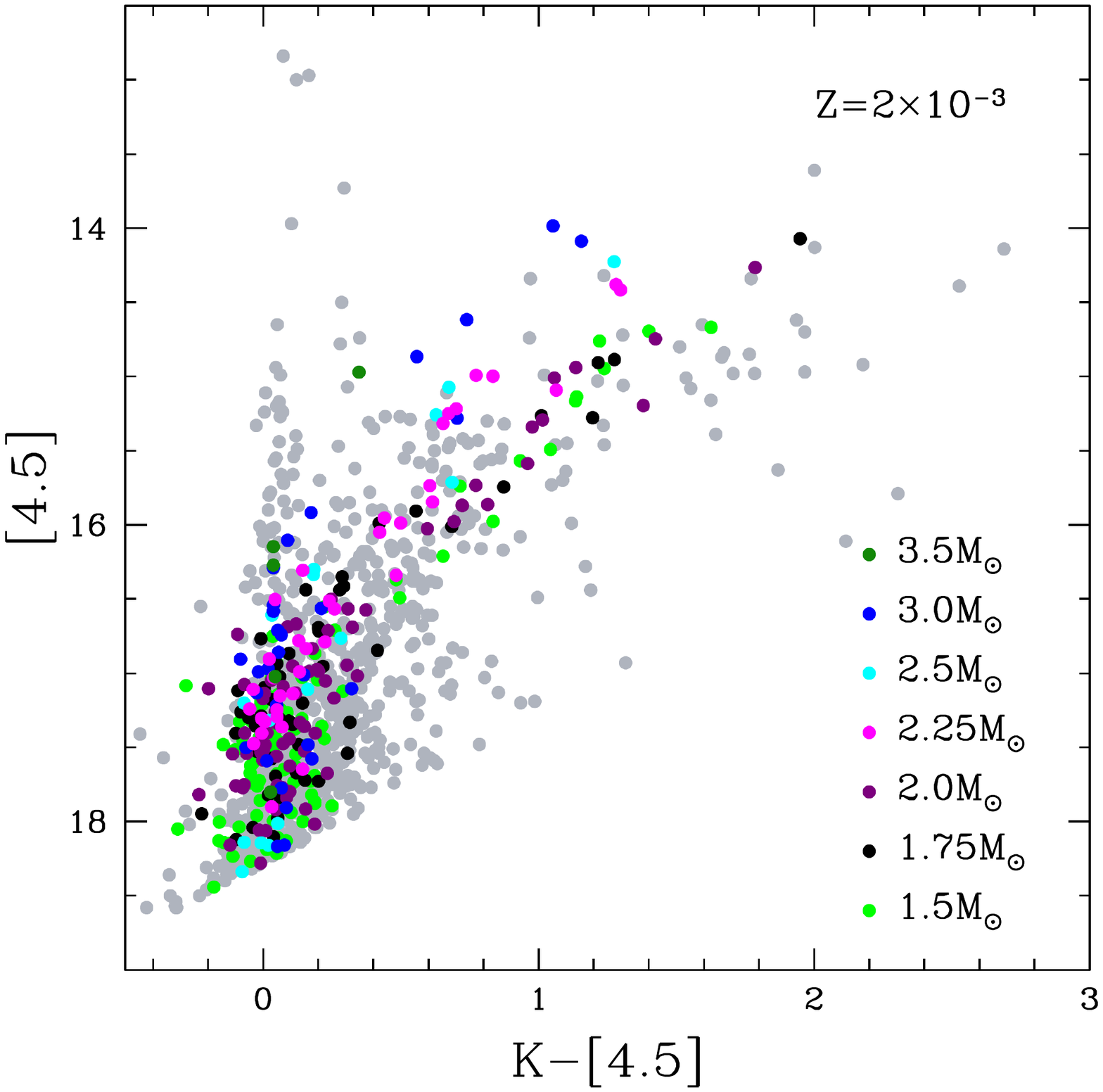}}
\end{minipage}
\begin{minipage}{0.33\textwidth}
\resizebox{1.\hsize}{!}{\includegraphics{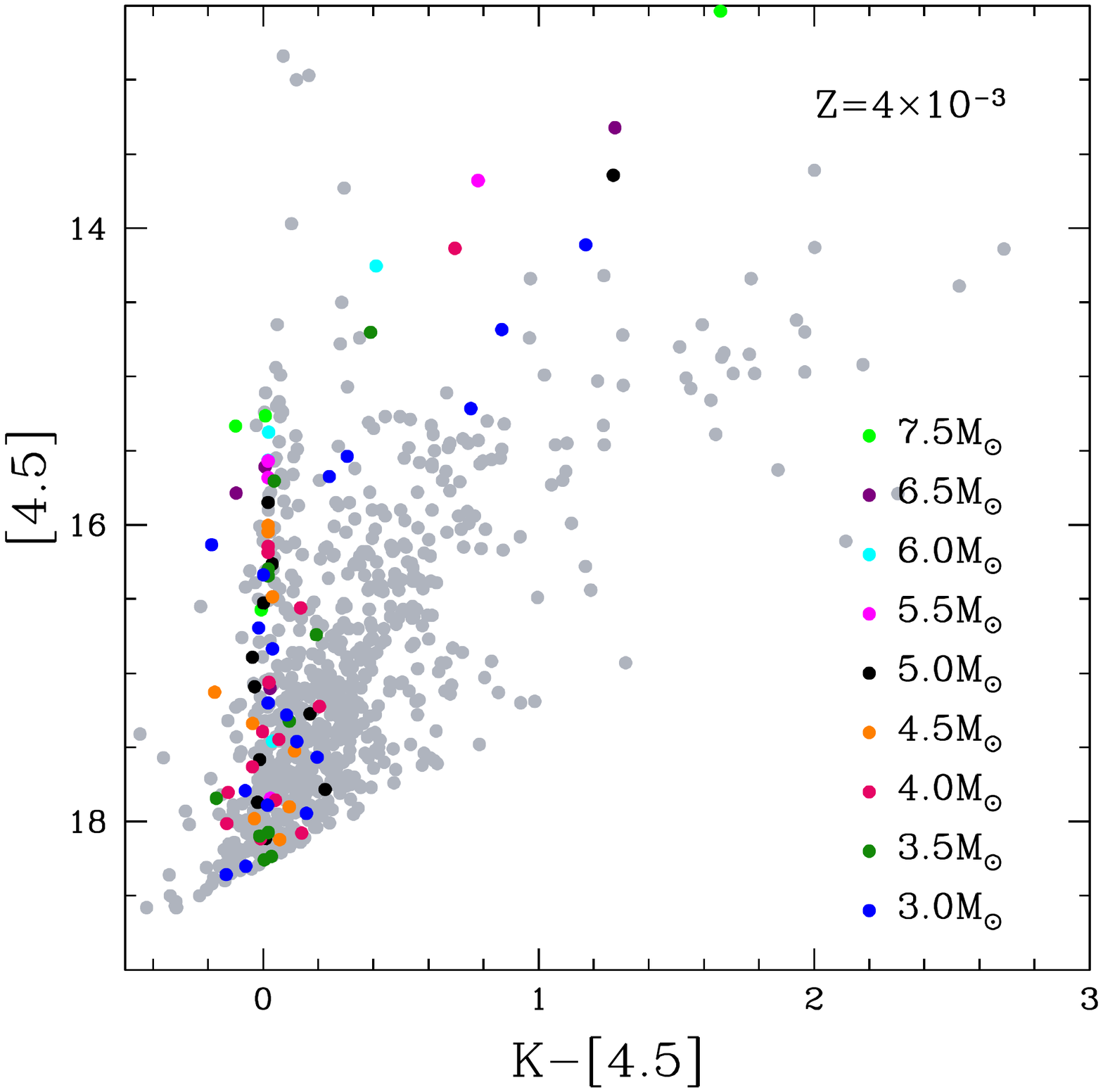}}
\end{minipage}
\vskip-30pt
\caption{The colour--magnitude ($K - [4.5], [4.5]$) diagram of the observed AGB candidate sample (grey) and the synthetic population with colour code indicating the initial mass; each panel refers to one of the three metallicities considered: $Z=10^{-3}$ (left), $Z=2\times 10^{-3}$ (middle) and $Z=4\times 10^{-3}$ (right).}
\label{CMD_sinmass}
\end{figure*}

Fig. \ref{CMD} shows the comparison between the observed (grey points)
and expected distribution of AGB stars in IC 1613, in the
CMD. The synthetic population in the left panel
is split among stars with a surface ratio $C/O<1$ (blue points) and
C--stars (red points). In the right panel the synthetic sample is
divided among the three metallicities used in the present analysis.
The three panels of Fig. \ref{CMD_sinmass} show the comparison between
the observations and the synthetic population in the CMD, for each of
the three metallicities considered; the extracted points are
colour--coded according to the mass of the progenitor.
Fig. \ref{CMD_CO} shows the comparison between the C-- and O--rich
samples classified by \citet{sibbons15} and the synthetic AGB
population.  We now discuss separately our interpretation of the
different groups of stars present in the observed sample.

\subsection{Oxygen--rich stars}
\label{oxygen}
According to our analysis, oxygen--rich stars constitute the majority
($\sim65\%$) of the AGB sample of IC 1613. 85\% of the oxygen--rich
AGB stars descend from low--mass stars, with initial masses $M\leq
3M_{\odot}$. These stars are evolving through the early-- and initial
TP--AGB phases, before reaching the C--star stage. In the CMD, these
objects populate the bluer region, with $-0.5<K-[4.5]<0.5$ and
$[4.5]>16.5$ mag. The circumstellar envelope of these stars is
essentially dust--free. Based on the results shown in
Fig. \ref{CMD_sinmass}, we deduce that the blue, low--luminosity
region in the CMD is mainly populated by metal-poor objects ($Z \leq
2\times 10^{-3}$), with masses $M\sim 1-1.5M_{\odot}$. The reason why these objects are the dominant population of the O--rich sample is twofold: a) lower mass stars ($M<1M_{\odot}$) make only 13\% of the oxygen--rich AGB stars because they are expected to evolve at magnitudes above the cut only in the very final thermal pulses, thus for a limited fraction of their AGB phase; b) higher mass ($M\sim 2-3M_{\odot}$) objects evolve above the threshold magnitude since the early--AGB; however, the number of these stars in the O--rich phase is smaller than their lower mass counterparts because any realistic mass function peaks towards lower masses and, more important, $2-3 M_{\odot}$ stars evolve as C--stars for a significant fraction ($\sim 50\%$) of their TP--AGB life.

From Fig.\ref{CMD_sinmass} and \ref{CMD_CO}, we note that 15\%
 of the stars populating the bottom--left region of the CMD with
 colours $K-[4.5] < 0.5$~mag and $[4.5]>16.5$ mag, have just reached the
 C--star stage; the small carbon excess with respect to oxygen is
the reason for the low degree of obscuration and their relatively blue
colours.

More massive stars ($M>3M_{\odot}$) account only for $15\%$ of the
O--AGB population in IC 1613. Though small in number, these stars can
be easily identified, because they are the only objects that evolve to
magnitudes brighter than $[4.5]=16.5$ mag, at $K-[4.5]\sim0$~mag. This is
clearly shown in Fig. \ref{CMD_tracks}, showing that the evolutionary
tracks of models of mass above $\sim 3M_{\odot}$ in the initial
TP--AGB phases reproduce the vertical finger in the CMD at
$K-[4.5]\sim0$~mag, extending to $[4.5]\sim15$ mag. In more advanced
phases, the ignition of HBB and the consequent increase in the
mass-loss rate favour the production of large quantities of silicates
\citep{paperIV}, with the gradual shift of the spectral energy
distribution towards longer wavelengths and redder $K-[4.5]$
colours. During this phase with efficient HBB these massive TP--AGB
stars are surrounded by silicates grains, with typical dimension
$a_{sil}\sim0.08 \mu m$ and optical depth $\tau_{10}\sim0.5$.  These
stars formed in recent epochs ($t<300$ Myrs ago) and, according to the
AMR by \citet{skillman14}, they belong to the most metal--rich
component, namely $Z=4\times 10^{-3}$. The relatively low number of
these objects stems from the small values of the IMF in this range of
masses and the short duration of their AGB phase. We note that in
order to reproduce the magnitude distribution of the stars in the
vertical finger of the CMD, it was necessary to increase the SFR by
\citet{skillman14} by a factor of 2 in the epochs between 100 Myr and
500 Myr ago.This step was necessary to increase by $\sim30\%$ the number of stars of mass $3M_{\odot}\leq M\leq 5.5M_{\odot}$, which populate the finger.  This assumption would be relaxed by taking into
account the partial contamination\footnote{Note that foreground contamination has already been taken into account and it is negligible ($<$1\%, Section 2).} from red super giant (RSG) stars,
expected to populate the brightest region of the CMD. In particular,
the stars identified by \citet{menzies15} as RSG, on the basis of their
small variability (open squares in Fig. \ref{CMD}), populate the
region $K-[4.5]\leq 0.5$~mag and $[4.5]<16$ mag. 
of the RSG stars contamination is difficult (out of the scope of this
paper) due to the poor resolution of the SFR in the most recent epochs
($t<500$ Myrs).

The left panel of Fig. \ref{CMD_CO} outlines a nice agreement between
the expected distribution of oxygen--rich stars in the CMD and the
sources classified as O--rich by \citet{sibbons15} (see
Sect. \ref{sample}). Some differences appear for the brightest [4.5]
magnitudes. Three, very bright stars, classified as oxygen--rich AGB
stars by \citet{sibbons15}, populate the region of the CMD
$(K-[4.5],[4.5])\sim(0.1,13)$, not covered by any of the evolutionary
tracks shown in Fig.\ref{CMD_tracks}. Such a large $[4.5]$ magnitudes
are indeed reached by massive TP--AGB stars ($5-7 M_{\odot}$), when
the HBB experienced is strong. However, during these phases silicate
dust is formed in significant quantities \citep{paperIV}, thus
producing a large IR emission. The evolutionary tracks (see
Fig. \ref{CMD_tracks}) then move to colours $K-[4.5]\sim1$~mag,
significantly redder than the values observed. For this reason we rule
out that these are AGB stars, rather we suggest that they belong to
the RSG sample of IC 1613. This result is in agreement with results
from the literature. Two out of the three bright stars
(J010501.65+020839.2 and J010458.36+020908.4) were classified as super
giants by \citet{menzies15}, as shown in Fig. \ref{CMD}. In the left panel of Fig. \ref{CMD_CO}, we also note that the observed O-rich sequence extends a bit redwards than our synthetic distribution, at colours $K-[4.5]\sim 0.4-0.5$. This is partly due to the criterion adopted to divide the C-rich and the O-rich stars in IC1613. An additional motivation could be that the distribution of the stars in this region of the CMD is extremely sensitive to the details of the description of the transition from the O-rich to the C-star phase, which is affected by some degree of uncertainty.

In the right panel of Fig.\ref{CMD} we see that the region of the CMD $-0.5<K-[4.5]<0.5$, where almost the entire oxygen--rich AGB population of IC 1613 is located, harbours stars with different metallicities. In particular, the [4.5] magnitude can be partly used as a metallicity discriminator: while the $[4.5]>17$ mag region is populated by O--rich AGB stars of any metallicity, only $Z=2,4\times10^{-3}$ objects evolve to magnitudes $16$ mag $<[4.5]<17$ mag, while the $[4.5]<16$ mag zone is populated exclusively by $Z=4\times10^{-3}$ AGB stars. This is a natural consequence of the AMR by \citet{skillman14}, according to which only $Z=4\times10^{-3}$ stars formed in the last 500 Myr, thus implying that only the most metal--rich ($Z=4\times10^{-3}$), AGB stars descend from most massive ($M>2.5M_{\odot}$) objects.

\subsection{Carbon stars}
\label{carbon}
The region of the CMD with colours $K-[4.5]>0.5$~mag is populated only by stars that have reached the carbon star stage\footnote{A few, very bright, oxygen--rich stars are also present in this region of the diagram.}; this is clearly shown in the distribution of C--stars, shown in the left panel of Fig. \ref{CMD}. The C--star sequence defines a diagonal band, crossing the CMD from the region $K-[4.5]\sim0.5$~mag to the reddest objects, with $K-[4.5]\sim2$. A few stars are found with $K-[4.5]\sim4$~mag. Following the analysis by \citet{flavia14}, we identify this diagonal band as an obscuration sequence, in turn related to the amount of carbon accumulated in the surface layers by repeated TDU episodes. In the phases immediately following the achievement of the C--star stage, stars produce a negligible quantity of dust ($\tau_{10}$ below $10^{-7}$). In these early phases, the C--star population partly overlaps with
the O--rich AGB stars in the CMD, as shown in Fig. \ref{CMD}. In the
subsequent evolutionary phases, more and more carbon is accumulated to
the external regions of the star, which favors the formation of
considerable quantities of dust, and the evolutionary tracks gradually
move redwards.

According to our analysis, the C--stars in IC 1613 descend from stars with initial mass in the range $1M_{\odot}\leq M\leq 3M_{\odot}$, corresponding to ages from 300Myr to 5.5 Gyr. 
The majority of these objects (see the histogram shown in the right panel of Fig. \ref{isto_mass} and the left panel of Fig. \ref{CMD_sinmass}) are the progeny of 1.25$M_{\odot}$ stars with metallicity $Z=10^{-3}$, formed 2.5 Gyr ago. A significant contribution ($\sim30\%$) to the C--star sample is provided by stars of initial mass $\sim1.5-2.5M_{\odot}$ and metallicity $Z=2\times10^{-3}$. Both components populate the region of the diagram at $K-[4.5]>0.5$~mag as shown by the filled histogram in the right panel of Fig. \ref{isto_mass}.

According to the recent analysis by \citet{ventura16}, the redwards
extension of the C--star sequence in the CMD, built with the IR bands,
is a valuable indicator of the progenitors of the stars belonging to
the observed sample, of the carbon accumulated in the external stellar regions,
and of the rate at which mass loss occurs. The analysis by \citet{ventura16} must be considered on a qualitative grounds, as it does not account for binarity, which might potentially increase the mass loss rate substantially, via non-spherical mass loss. However, as we will clarify in the following, a similar, tight correlation between the degree of obscuration (hence, the $K-[4.5]$ colour) and the parameters given above, cannot be applied in this case.
In the case of IC 1613, the
right panel of Fig. \ref{CMD} and the left and middle panels of
Fig. \ref{CMD_sinmass} show that both $Z=10^{-3}$ stars of initial
mass $\sim1.0-1.25 M_{\odot}$ and their $Z=2\times10^{-3}$
counterparts, with initial mass $\sim1.5-2.0M_{\odot}$, populate the
region of the CMD at $K-[4.5]\sim2$~mag. In the MCs,
\citet{ventura16} identified a unique class of progenitors for stars
in the reddest side of the C--star sequence, but here we find a
miscellaneous distribution of masses and chemical compositions. The
reason for this is that the SFH of IC 1613 is rather constant
\citep{skillman14}, whereas the SFH of the MCs present peaks of
intense star formation \citep{harris09}, during which the formation of
the stars currently populating the red side of the C--star sequence
occurred \citep{ventura16}.

In the analysis of the C--stars populating the reddest regions of the
CMD, we note that objects with $Z=10^{-3}$ that have the reddest
$K-[4.5]$ colours are those evolving through the very final TP--AGB
phase, when the envelope is about to be entirely lost. This is deduced
by comparing the terminal point of the $1M_{\odot}$ track, in the left
panel of Fig \ref{CMD_tracks}, with position of the stars in the CMD
at $K-[4.5]\sim2$~mag. Conversely, for the C--star population of higher
mass and metallicity, our models predict that the very final TP--AGB phases are at much redder
colours than the AGB stars considered in the CMD\footnote{The presence of extremely obscured stars, not detected in the K band, can not be neglected.} (Fig.\ref{CMD_tracks}). This
difference is motivated by the greater amounts of carbon accumulated
at the surface of these objects compared to their counterparts of
lower mass, which favour a significant increase in the mass--loss rate
thus provoking a considerable shortening of the evolutionary
time--scale. It is interesting to note that, owing to the larger initial masses involved, the $Z=2\times 10^{-3}$ stars evolve to brighter $[4.5]$ magnitudes, compared to the $Z=10^{-3}$ objects (see right panel of Fig. \ref{CMD}). Our models predict that the stars on the reddest side of
the C--star sequence are surrounded by carbonaceous particles, of size
$0.1-0.15 \mu m$, with optical depth $\tau_{10}\sim 0.1$. Smaller dust
grains ($\sim0.05 \mu m$) of SiC are also expected to be present in
the circumstellar envelope of this group of stars.
In summary, owing to the combination of the SFH and the AMR, which make the mass distribution to peak at low-metallicity, $\sim1-1.5M_{\odot}$ stars (see Fig. \ref{isto_mass}), we find that the majority of the most obscured AGB stars, giving the largest contribution to dust production in IC1613, belongs to the lower-metallicity stellar component, with $Z\leq 2\times10^{-3}$.

\begin{figure}
\resizebox{1.\hsize}{!}{\includegraphics{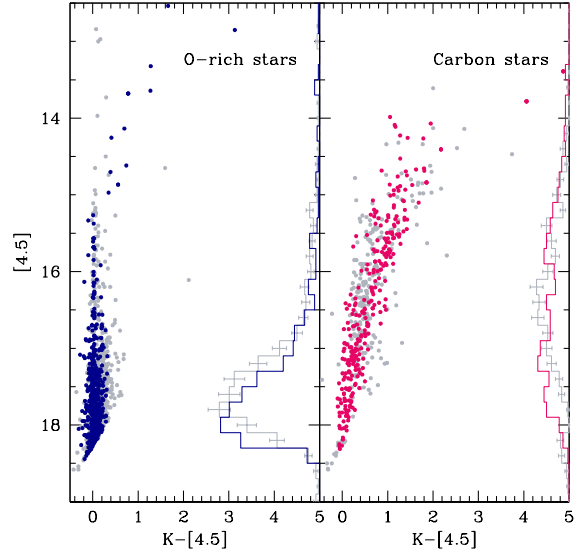}}
\vskip-50pt
\caption{Left panel: the colour--magnitude ($K - [4.5], [4.5]$) diagram of the O--rich stars (gray points) classified by \citet{sibbons15} (see section \ref{sample}) and O--rich stars (blue) from synthetic modelling. Right panel: the same distribution for C--AGB candidates (gray points) classified by \citet{sibbons15} (see section \ref{sample}) and carbon stars (red) from the synthetic population.}
\label{CMD_CO}
\end{figure}

There is nice agreement between the objects classified as C--stars by
\citet{sibbons15} and the stars with $C/O>1$ in the simulated
population. This can be seen in the right panel of Fig. \ref{CMD_CO},
which shows that the color-magnitude distribution of the observed
sample is reproduced by the models. Only in the lowest part of the
diagram do we find a mismatch between the two samples, where a
selection effect limits the number of C stars identified
photometrically.

The histogram shown in the left panel of Fig. \ref{CMD} outlines a
slight mismatch between the expected and observed number of stars in
the magnitude range $16$ mag $< [4.5] <16.5$ mag. According to our
modelling, only a small number of AGB stars should be present in that region of
the CMD, thus suggesting a too-fast evolution through that magnitude
interval. That part of the CMD is populated by $1.5-2M_{\odot}$
objects that have just achieved the C--star stage. The small number of
predicted stars is triggered by the sudden shift of the SED to longer
wavelengths as the C--star phase begins, which in turn, is motivated
by an efficient dust production mechanism. This suggests the need for
more detailed treatment of dust formation in the cases when the
radiation pressure is not highly efficient in accelerating the wind.
This is not surprising, considering that these are the only situations
where the results obtained are significantly dependent on the adopted boundary
conditions, particularly the velocity ($v_0$) with which the
wind enters the condensation region. We confirmed this by changing $v_0$
from 1km/s to 2 km/s, which leads to a slower evolution through that region
of the CMD with $K-[4.5]<1$~mag. Note that this assumption has a
negligible effect on stars populating reddest region of the diagram,
where the dust production sufficiently accelerates the wind, eliminating
any dependency on the initial velocity. This choice leads to a much
better agreement between the observations and the expected
distribution of stars.

\begin{figure}
\resizebox{1.\hsize}{!}{\includegraphics{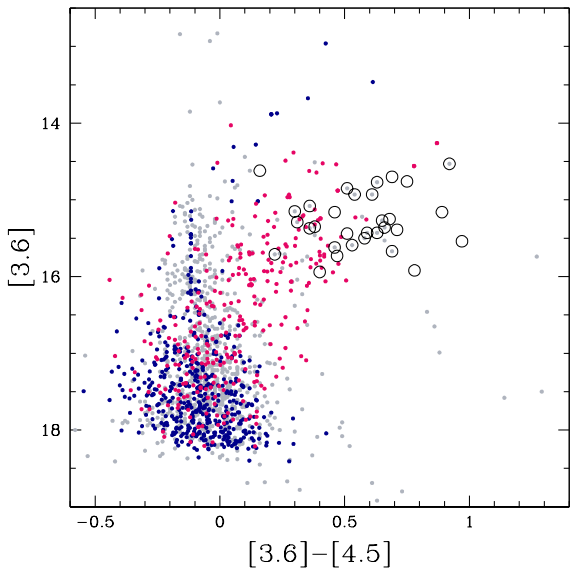}}
\vskip-50pt
\caption{The colour--magnitude ($[3.6] - [4.5], [3.6]$) diagram of the sample of xAGB stars classified by \citet{boyer15II} (open circles); the blue and red dots indicate respectively the O--rich and carbon stars from synthetic modelling. The full AGB candidate sample (grey points) is also shown. Empty open circles refer to stars not selected as AGB candidates by \citet{sibbons15}.}
\label{CMD_xagb}
\end{figure}

\subsection{xAGB stars}
\label{xAGB}

Recent studies of the MCs outlined the presence of extremely red
objects, with $(J-K)>2$~mag; spectroscopic analysis showed that these
AGB stars are mainly carbon stars (Woods et al. 2011, Boyer et al. 2015c,
Dell'Agli et al. 2015a,b).

\citet{blum06} classified these objects as ``extreme--AGB" (xAGB); a
straight application of that classification cannot be adopted here
since the $J$-band magnitude is not available for a large fraction of
AGB stars in IC 1613.  For this reason, \citet{boyer15II} identified
xAGB stars in this galaxy using both the variability index (with 2
epochs) introduced by \citet{vijh09} and an IR colour--magnitude
classification ($[3.6]-[4.5] > 0.1$ mag and $M_{3.6} = -8$ mag). Based
on this criterion, they identified 30 xAGB stars in IC 1613, shown as
open circles in Fig. \ref{CMD_xagb}. In the same ([3.6]-[4.5],[3.6])
diagram, we report the observed stars of our catalogue (grey points)
and the results from our simulation, divided among carbon stars (red)
and oxygen--rich objects (blue).  The
xAGB sample of IC 1613 are modeled to be the progeny of 1--1.25$M_{\odot}$ stars of
metallicity $Z=10^{-3}$ and $1.5-2.5M_{\odot}$ objects with
$Z=2\times10^{-3}$. These stars are currently evolving through the
final TP--AGB phase, after several TDU events have determined the
transition to the C--star stage. All these sources are modeled to be surrounded by
carbon dust, with optical depths in the range $0.01<\tau_{10}<0.1$. A
tighter identification of the mass and chemical composition of the
individual xAGB is not straightforward; the only exceptions are the
sources located at $[3.6]-[4.5]>0.5$~mag, which descend from $\sim
2M_{\odot}$ stars of metallicity $Z=2\times10^{-3}$.

This interpretation of the xAGB stars is also confirmed by their position in the ($K-[4.5],[4.5]$) diagram, where they populate the carbon star sequence (see Section \ref{carbon}) up to $K-[4.5]\sim4$~mag. Note that the reddest xAGB stars (empty open circles with $K-[4.5]>2$~mag in Fig. \ref{CMD}, left panel) are not classified as AGB candidates by \citet{sibbons15}, mainly because they are obscured and thus not detected in the J band. Therefore, we can conclude that: a) the criterion assumed by \citet{sibbons15} entails a probable loss of the most obscured stars; and b) the colour extension of the carbon star sequence up to $K-[4.5]\sim4$~mag predicted by the models is confirmed by the observations.

\begin{table*}
\begin{center}
\caption{Properties of the synthetic AGB population} 
\label{tab_dpr}
\begin{tabular}{lcccccc}
\hline
&$M/M_{\odot}$ & $Z$ & age & \% & $a$ [$\mu m$] & DPR $[M_{\odot}/yr]$\\ 
\hline
O--rich&&&&&&\\
low-mass& $\leq3M_{\odot}$& $\leq2\times10^{-3}$ & 0.3 - 12 Gyr & 55\% & $\sim0.001$ & $\sim10^{-8}$ \\
high-mass& $>3M_{\odot}$& $4\times10^{-3}$ & 40 - 200 Myr & 10\% &  0.007-0.08 & $\sim9\times10^{-8}$ \\
&&&&&&\\
C--rich& $1-3M_{\odot}$& $\leq2\times10^{-3}$ & 0.3-5.5 Gyr & 35\% & 0.003-0.18 & $\sim5\times10^{-7}$ \\
\hline
\end{tabular}
\end{center}
\end{table*}

\subsection{Dust--production rate}

The estimation of the dust-production rate (DPR) from AGB stars is
still a matter of debate. Several observational methods are adopted,
usually based on the IR excess and the estimated mass-loss rate. This
was the case for the MCs, where the DPR was determined either by
excess in the {\it Spitzer} 8 $\mu m$ band \citep{srinivasan09,boyer12}
or by the colour excess \citep{matsuura09,matsuura13}. An alternative approach
fit the SED of the individual objects
\citep{riebel12,srinivasan16}. These kind of estimates become very
difficult in other dwarf galaxies because the larger distances and the
lower metallicities make the estimation more uncertain, particularly
without the support of theoretical models able to describe the
dust-production process in metal-poor environments. The first attempts to
estimate the DPR based on the theoretical description of the dusty
circumstellar envelope of AGB stars (in the MCs) were published by
\citet{zhukovska13,schneider14,flavia15a}.

To reach this goal in the case of IC 1613, we followed the same
approach used for the LMC (see Sect. 8 in \citet{flavia15a} for more
details): at each time step of the AGB evolution, the DPR
($\dot{M}_d$) of the individual stars is calculated as the product of
the mass--loss rate and the degree of condensation of each dust
species considered. For oxygen--rich stars, we consider the percentage
of gaseous silicon and aluminum condensed, respectively, to silicates
and alumina dust. For carbon stars we take into account the
fraction of carbon and silicon condensed into solid carbon and SiC
particles.

Our predicted contribution to the total DPR of IC 1613 from each class
of stars is summarized in Table \ref{tab_dpr}. O--rich AGB stars
contribute $\dot{M}\sim10^{-7}M_{\odot}/yr$, with a dominant
contribution from silicate dust (90\%) compared to alumina dust
($\sim10\%$). The O--rich stars provide most (above $\sim 90\%$) of
the DPR from stars of mass above $\sim3M_{\odot}$; although limited in
number, these objects produce much higher quantities of dust than
their counterparts of smaller mass. The latter are evolving through
the O--rich phase previous to the formation of a carbon star, when
limited dust production occurs. The carbon star contribution to the
DPR is $\dot{M}_d\sim5\times10^{-7}M_\odot/{\rm yr}$. Most of the dust
formed is solid carbon, with a very modest (below $\sim3\%$)
contribution from silicon carbide; this is due to the small
metallicities and the resulting scarcity of gaseous silicon in the
wind.  Among the various stars belonging to the C--rich group (see
discussion in section \ref{carbon}), those providing most of the DPR
($\sim80\%$) are the progeny of $2M_{\odot}$ objects of metallicity
$Z=2\times10^{-3}$; the remaining contribution ($\sim20\%$) is given
by $Z=10^{-3}$ stars of initial mass $\sim1M_{\odot}$. The estimate of DPR from C-stars is affected by the uncertainties on the rate of mass loss, the efficiency of the TDU (determining the carbon available in the surface regions), the details of condensation of carbon molecules and the SFH. We estimate that the overall uncertainty associated to the given DPR is below $\sim 30\%$.

The C-rich population is giving the dominant contribution also to the gas pollution of IC1613. According to our analysis, the mass of carbon gas ejected by AGB stars into the ISM is $\dot{M}^{C}_{g}\sim2.7\times10^{-6}M_{\odot}/yr$, against the corresponding quantities of oxygen gas, estimated to be $\dot{M}^{O}_{g}\sim1.1\times10^{-6}M_{\odot}/yr$. This reflects the higher number of carbon stars currently loosing mass at larger rates compared to the O-rich population. However, we expect that the current C/O ratio of the ISM is lower, as our estimate is based only on the AGB winds, thus neglecting additional contributions from, e.g., the pristine gas in the ISM and the mass lost by supergiants. The gas-to-dust ratio from the AGB population is $\dot{M}_g/M\dot{M}_d \sim 1000$, estimated from the gas and dust mass-loss rates from the synthetic AGB population. This value is in agreement with the expectation that gas-to-dust ratios increase in lower metallicity environments (e.g., van Loon et al. 2000).

\section{Conclusions}
\label{conclusions}
We study the population of AGB stars in the dwarf galaxy IC 1613
comparing models to the combination of near--IR photometry from
\citet{sibbons15} and mid--IR data from the DUSTiNGS catalogue
\citep{boyer15I}. To this aim, we follow the same approach used to
investigate the evolved stars in the MCs, based on a population
synthesis method; wherein the observed distribution of stars in the
colour--magnitude plane is compared to the simulated population. We
build the simulations with the latest generation of AGB models, which
include dust formation in the circumstellar envelope. To date, this is
the first study where these models are applied to a galaxy external to
the MCs system. The nice agreement between the models and the
observations allows an interesting characterization of the individual
sources observed, with the determination of the mass, chemical
composition, and progenitor formation epoch. We find that the AGB
sample of IC 1613 is composed of 65\% oxygen--rich objects and 35\%
carbon stars. The first group is dominated by a low--mass
($M<2M_{\odot}$) component, accounting for 55\% of the overall AGB
population, formed 2--5Gyr ago, with metallicity below
$Z\sim2\times10^{-3}$. These stars are evolving through the AGB phases
previous to the achievement of the carbon star stage and their
environment is essentially dust--free. The remaining oxygen--rich
stars comprise a younger population (ages below 200 Myr) of higher
metallicity ($Z=4\times10^{-3}$) stars, descending from objects with
mass above $3M_{\odot}$. This young population can be easily
identified in the colour--magnitude ($K-[4.5],[4.5]$) diagram, as they
populate a vertical finger at $K-[4.5]\sim 0$~mag and $[4.5]<16.5$ mag.
The carbon stars in IC 1613 descend from stars in the range
1--2.5$M_{\odot}$ and metallicities $Z\leq2\times10^{-3}$, formed
between 300 Myr and 5.5 Gyr ago. Their IR properties are mainly
determined by the amount of carbon dust in their circumstellar
envelope. Similarly to the MCs, these objects trace an obscuration
sequence in the CMD that extends to $K-[4.5]\sim4$~mag. Our models predict
that the sources exhibiting the largest degree of obscuration are
surrounded by solid carbon particles, with size $0.1-0.15\mu m$. These
stars give the dominant contribution ($\sim85\%$) to the overall
dust--production rate of IC 1613, predicted to be
$5\times10^{-7}M_{\odot}/yr$ by our simulations.

The present work not only represents a major step in the understanding
of the gas and dust pollution from AGB stars, but also demonstrates
that these models can be used to characterise evolved stellar
populations in diverse environments.  These findings are timely with
the upcoming instruments such as the European Extremely Large
Telescope or James Webb Space Telescope, which will enormously
increase the number of galaxies where AGB stars will be
resolved. Analyses similar to this will provide important information
about star formation histories, which are currently inferred only from
integrated spectral energy distributions.

\section*{Acknowledgments}
The authors are indebted to the anonymous referee for the careful reading of the manuscript and for the several comments, that help improving significantly the quality of this work.
F.D.A. acknowledges support from the Observatory of Rome. M.D.C. acknowledges Adriano Fontana and the contribution of the FP7 SPACE project ASTRODEEP (Ref.No:312725) supported by the European Commission. D.A.G.H. was funded by the Ram\'on y Cajal fellowship number RYC$-$2013$-$14182 and he acknowledges support provided by the Spanish Ministry of Economy and Competitiveness (MINECO) under grant AYA$-$2014$-$58082-P..

\end{document}